\documentstyle[12pt,epsf]{article}
\begin{document}
\begin{center}
{\ {\Large {\bf Asymmetrical and Non-equilibrium Expansion }} \\
   {\small \  } \\
   {\Large {\bf  of Pion Gas  }} \\
   {\small \ } \\
   {\large {\bf Created in Ultrarelativistic Heavy Ion Collision }}}

\vskip 0.75cm

{\em Peter Filip} \footnote{e-mail: filip@savba.sk} \\
\vskip 0.35cm
{\it Institute of Physics, Slovak Academy of Sciences, SK 84228 \\
\vskip 0.15cm
Faculty of Mathematics and Physics, Comenius University, SK 84215 \\
\vskip 0.15cm
Bratislava, Slovak Republic} \\

\vspace{0.32cm}
\end{center}

\begin{abstract}
Results of the computer simulation of dilepton production
from the expanding pion gas
created in ultrarelativistic Pb+Pb 160 GeV/n collisions are presented.
Finite size of the expanding pion gas influences invariant
mass spectrum of dileptons. Sensitivity of the shape of dilepton mass spectrum
to the initial stage of the pion gas is discussed. Second order asymmetry
in azimuthal distribution of dileptons is predicted for non-central collisions.
Non-equilibrium nature of the pion gas expansion is found to exhibit
itself in rapidity distribution of produced dilepton pairs.
\end{abstract}


\section{Introduction}\label{sec1}
Goal of heavy ion collision (HIC) experiments is to reveal properties
of the compressed hot nuclear matter created during the collision of
heavy nuclei. Final momentum distributions of most abundant
particles - hadrons are however influenced (if not determined)
during the dilute and late freeze-out stage of the heavy ion collision.
Interesting information about the early stage of the collision is in this case
hidden by subsequent collective effects of strongly interacting
hadrons.

Fortunately other particles - leptons and photons
produced in all stages of HIC process, are not influenced by
the later hadronic phase. Thus distributions of leptons and photons
can provide us with more or less direct information about the early
stages of the collision process.

Recent experimental results on invariant
mass distribution of dilepton pairs \cite{NA50_A610_331,CERES} have
attracted increased attention due to the observed excess of dilepton
yield at regions $1.5 < M_{\mu\mu} < 2.5$GeV (NA50) and
$0.2 < M_{ee} < 1.5$ GeV (CERES).
Theoretical analysis on dilepton pair production \cite{Shuryak}
leads to the conclusion that unconventional production mechanisms
(e.g. modified $\rho $-meson mass \cite{Shur_7,Song_Nakano} or
in-medium pion dispersion relation \cite{Fuchs,Shur_13}) are necessary to
explain present experimental data.

In this contribution specific results of the computer simulation of dilepton
pairs production from expanding pion gas are presented. Besides the
invariant mass distribution of the produced dileptons also other
properties of the dileptonic signal produced from the pion gas via
$\pi^+\pi^-$ annihilation channel are analyzed.

Short description
of the computer simulation is given in next section.
Influence of the finite size
of pion gas (and its initial state) on invariant mass
spectrum of dileptons is discussed in Section 3 using results of the
computer simulation. In Section 4 rapidity distribution of dilepton pairs
obtained from the simulation is presented. Non-equilibrium
features of the pion gas expansion are discussed in this context.
Second-order azimuthal asymmetry in the
transversal momentum distribution of dileptons is predicted by the
simulation of non-central Pb+Pb 160 GeV/n collisions in Section 5.
Possible consequences of the effects
found and presented here on other areas of HIC research are scetched 
at the end of this paper.

\section{Simulation of Dilepton Production}\label{secx}
For the simulation of dilepton production from the expanding
pion gas rescattering model \cite{APS96,APS97}
(inspired by the model of T.J.Humanic \cite{Hum_Pis}) was used.
Initial stage of pion gas in Pb+Pb 160 GeV/n collisions
(positions, momenta and time of creation of pions) was taken from events
generated by cascade program \cite{Zavada}. In the rescattering program
pions move in small time steps $\Delta T=0.1 $fm as free relativistic
particles and their momenta are changed in mutual collisions.
In each time step distance of pions $d_{ij}=|\vec x_i - \vec x_j|$ is
compared to critical distance $d_c(s)=\sqrt{\sigma _{\pi}(s)/\pi}$,
where $\sigma _{\pi}(s)$ is total elastic $\pi\pi $ cross section.
If condition $d_{ij}<d_c(s)$ is fulfilled collision procedure for a
given pair of pions is called.
More detailed description
of the simulation can be found in \cite{APS97,nucl_th51}.

Two independent methods of the implementation of dilepton production
via $\pi^+\pi^-$ annihilation channel were used in the simulation.
First one is based on standard expression for the rate of dilepton
production \cite{Ruusk}:
\begin{equation}
\frac{dN_{ee}}{dM^2d^4x} =
\int \frac{d^3k_1}{(2\pi)^3}f(k_1,x)
\int \frac{d^3k_2}{(2\pi)^3}f(k_2,x)
[v_r \sigma_e(M^2) \cdot \delta(M^2-(k_1+k_2)^2)]
\label{ee1}
\end{equation}
where $\sigma_e(M)$ is the cross section for
$\pi ^+\pi ^- \longrightarrow e^+e^- $ at given $\sqrt{s}=M$,
$s=(k_1+k_2)^2$, $v_r$ is the relative velocity
$v_r=\sqrt{s(s-4m^2_{\pi})}/2E_1E_2$ and $f(k,x)$ is distribution function
of pions in phase space, $x=(\vec x,t)$. Distribution function $f(k,x)$
is calculated by rescattering program in our case. It depends mainly on the
isospin averaged cross section
$\sigma_{\pi}$ for  elastic $\pi \pi $
interaction which regulates scattering of pions and it is influenced also
by formation time $\tau_f$ and interaction time $t_i$
parameters of the rescattering model used \cite{APS97}.

After multiplying and dividing right-hand side of Eq.(\ref{ee1}) by
$\sigma _{\pi}(M^2)$ and integrating over $d^4x$ the rate of dileptons
from the pion gas can be expressed as \cite{nucl_th51}:
\begin{equation}
\frac{dN_{ee}}{dM^2} =
R\cdot \frac{\sigma_e(M^2)}{\sigma_{\pi}(M^2)}
\cdot \frac{dN_{\pi}}{dM^2}
\label{ee3}
\end{equation}
where $\frac{dN_{\pi}}{dM^2}$ is 
distribution of $\pi\pi$ collisions
as a function of invariant mass $M$ of the colliding $\pi\pi$ pairs
and $R$ denotes
ratio of the number of $\pi ^+\pi ^-$ collisions to all
$\pi \pi $ interactions.
Distribution $\frac{dN_{\pi}}{dM^2}$ is calculated by the rescattering program.

\vskip0.15cm
\centerline{\epsfxsize=10.7cm\epsffile{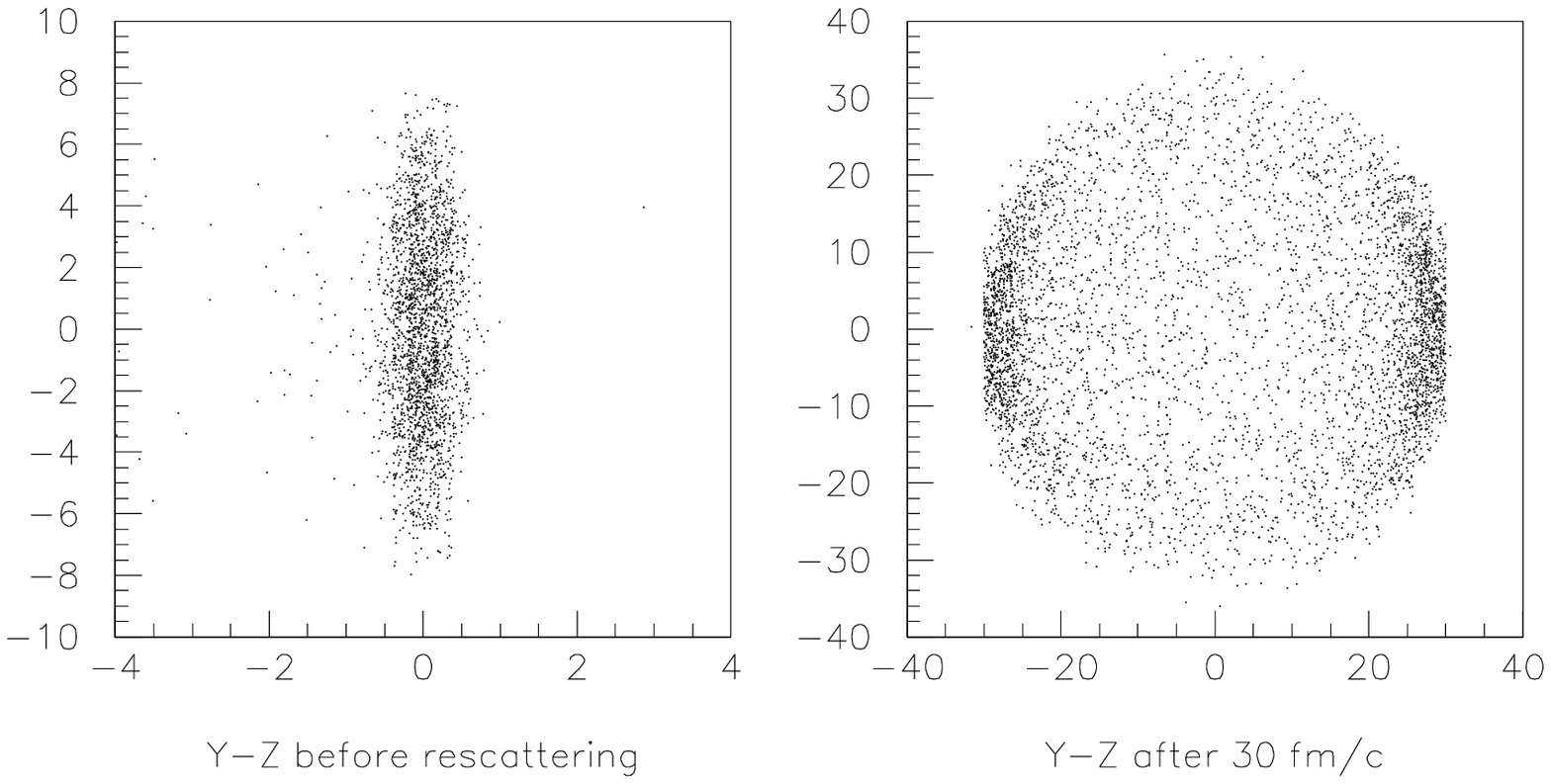}}
\vskip1.7pt
\centerline{\parbox{13cm} {\small {\bf Fig.1} Distribution of pions in 
the reaction plane before and after the rescattering process in CMS frame
of the collision. Axis values
are in {\it fm} units.
}}
\vskip0.15cm

At the same time another method for the simulation of $\pi^+\pi^-$ annihilation
in the expanding pion gas was used.
Cross section
$\sigma _e(s) $ for the process $\pi^+\pi^- \rightarrow e^+e^-$
was multiplied by factor 13000 in the rescattering program and pairs of pions
were tested (in each time step)
for the critical distance $d_c^e(s)$ calculated from
this enlarged cross section:
\begin{equation}
d_c^e(s)=\sqrt{13000\cdot \sigma _e(s)/\pi}> |\vec x_i -\vec x_j|
\label{Dce}
\end{equation}

When this condition is fulfilled for a given pair of pions
invariant mass $M$ and	momenta of pions $\vec p_i,\vec p_j$ are remembered
(stored into file).
Positions and momenta
of pions fulfilling condition (\ref{Dce}) are left unchanged in the program.
In this way probability for
the process of pion annihilation is enhanced to the level comparable with
the elastic pion-pion interactions. At the same time process of the
expansion of pion gas is not influenced by this high probability of
pion-pion annihilitaion because pions are not annihilated finally.
Just possibility of this process is recorded.

Results obtained by the above described two independent methods
of dilepton production simulation
were nearly identical in all features (invariant mass distribution,
$p_t$ distribution, azimuthal and rapidity distribution) tested so far.

\section{Invariant Mass Distribution}\label{sec2}

Invariant mass distribution of dilepton pairs produced in ultrarelativistic
heavy ion collisions can be divided into separate regions. Dileptons
with invariant mass $M>3$ GeV are produced mainly in the process
of parton-parton annihilation. Region of $M$ below 300 MeV is dominated
by Dalitz decays of $\pi ^0,\eta,\omega$ and $\eta '$ mesons. Also
electromagnetic bremsstrahlung contributes to this low mass region of
dilelectron pairs. Dilepton pairs at region of $M$ close to 1GeV are
decay products of $\rho$,$\omega $ and $\phi $ mesons.
In the region between 280 MeV and 1 GeV thermal production from $\pi^+\pi^-$
annihilation contributes. This list of dilepton production channels would
be incomplete if we did not mention dileptons from decays of $J/\psi $,
$\psi '$ and couples of $D\bar D$ mesons.

$\pi^+\pi^-$ annihilation channel of dilepton production plays important role
in the case of ultrarelativistic heavy ion collisions
due to high multiplicities of secondary produced pions (above 2000
in central Pb+Pb 160 GeV/n collisions).

\vskip0.25cm
\centerline{\epsfxsize=14cm\epsffile{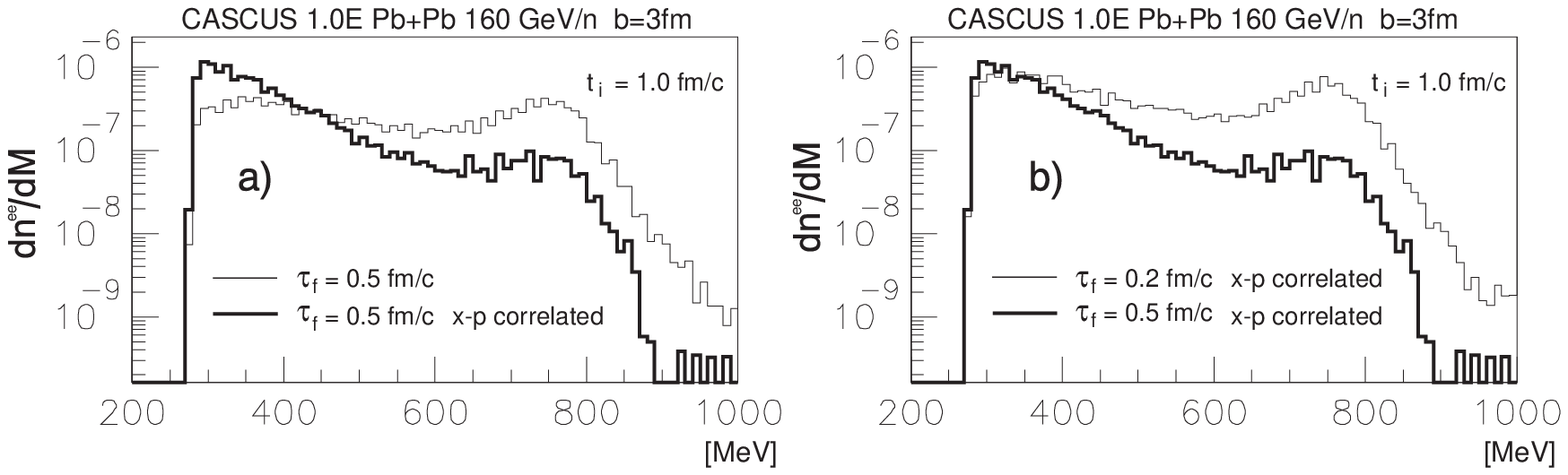}}
\vskip1.7pt
\centerline{\parbox{13cm} {\small {\bf Fig.2} 
Invariant mass distributions of dileptons obtained by the simulation.
}}
\vskip0.15cm

In this section invariant mass spectra of
dileptons originating from $\pi^+\pi^-$ annihilation generated by the
rescattering simulation are presented.
In the paper \cite{nucl_th51} mainly influence of formation time parameter
on the shape of dilepton spectrum was investigated.
Here dependence of the dilepton mass spectrum on the initial
state of the pion gas is studied. 

Initial momenta and positions of pions
in events produced by cascade generator \cite{Zavada} are uncorrelated.
It means
that momentum of the pion does not depend on coordinates of its
creation e.g. in transversal plane. Invariant mass spectrum obtained for
these $x-p$ uncorrelated events is shown
on Fig.2a).

Thick line on Fig.2a) is obtained for same values
of $\tau _f$ and $t_i$ parameters of the rescattering model but for
$x-p$ correlated events. These events were obtained by the following
modification of events produced by cascade generator \cite{Zavada}: 
For each pion scalar product of its transversal momentum and
position in transversal plane $\vec p_t \cdot \vec x_t$ was evaluated.
In the case of negative numerical value of the scalar product,
position of the pion
$\vec x_t$ was replaced by the oposit position in transversal plane
$-\vec x_t$. In this way momentum-space correlation was generated
in the initial state of pion gas. Momenta of pions (not changed) are oriented
predominantely out from the center of transversal plane.

Resulting invariant mass spectrum of dileptons calculated for $x-p$
correlated events differs significantly
from the spectrum of uncorrelated events.
Low mass dilepton pairs
are strongly enhanced in the case of $x-p$ correlated events. In Fig.2b
influence of formation time parameter $\tau_f$ on invariant mass spectrum
of dileptons is shown for $x-p$ correlated events.
Small values of formation time parameter
enhance production of dilepton pairs in $\rho$-meson region 
as it was found and discussed
paper \cite{nucl_th51}.

What is physical origin of the low mass dilepton
enhancement in the case of $x-p$ correlated events ? Why non-zero
values of formation time suppress dilepton pairs in $\rho$-mass region
compared to the low mass region pairs ?

Two independent effects originating from the finite size of
pion gas can serve as explanation. First one was already discussed
in the work \cite{nucl_th51}. This consideration is reproduced very
shortly here. Formation time \cite{Form_time} excludes pions from
the rescattering process and therefore also from the possibility of
annihilation into dilepton pairs. Faster pions are forbidden to
interact for longer time due to the Lorentz dilation of the formation
time implemented in the rescattering program. Due to the finite size of
the pion gas faster pions can escape from the interaction region
without any rescattering or annihilation. Thus invariant
mass spectrum if dileptons is enhanced at low mass region by interactions
of slow pions. 

At the same time another mechanism also resulting from finite size of the 
expanding pion gas influences invariant mass spectrum of dileptons. Invariant
mass quantity depends not only on the total size of momenta of pions being
scattered but also on the angle between the momenta:

\begin{equation}
M=\sqrt{(E_1+E_2)^2 -(\vec p_1 + \vec p_2)^2} =
  \sqrt{2E_1E_2 + 2m_{\pi}^2 - 2p_1p_2\cos(\phi )}
\label{Mfi}
\end{equation}

For small values of scattering angle $\phi $, it means for pions moving
in parallel directions, invariant mass quantitiy (\ref{Mfi}) is small.
At the same time for pions moving in oposit directions invariant
mass is large for the same values of momenta of pions. This
effect can play significant role if the mean free path of
interacting particles is comparable with the initial size of the expanding
gas. This seems to be the case of the pion gas created in ultrarelativistic
heavy ion collisions.

Momentum-space correlation leads to the higher probability of collisions
with small scattering angles and therefore it enhances production of low
mass dileptons (see Fig.2a).
In the case of ultrarelativistic heavy ion collisions momentum-space
correlation at the initial stage of pion gas
can be generated also by transversal flow in prehadronic phase.

\section{Rapidity Distribution}\label{sec3}
Rapidity distribution of dilepton pairs produced by pion annihilation
is determined by rapidity distribution of momentum sum
$\vec p_{\pi\pi}=\vec p_1 + \vec p_2 $ of anninhilating pions
($\vec p_1$,$\vec p_2$ are momenta of $\pi ^+$ and $\pi ^-$).
This quantity is conserved in the process of $\pi^+\pi^-$ annihilation,
$\vec p_{\pi\pi}=\vec p_{ee}$. In Fig.3a) and b)
rapidity distributions of pions before and after
the rescattering process are shown.
Rapidity distribution of pions is not influenced significantly
by the rescattering process. However resulting rapidity distribution of
dilepton pairs exhibits unexpected minimum at midrapidity surrounded
by two maxima.

\vskip0.25cm
\centerline{\epsfxsize=14cm\epsffile{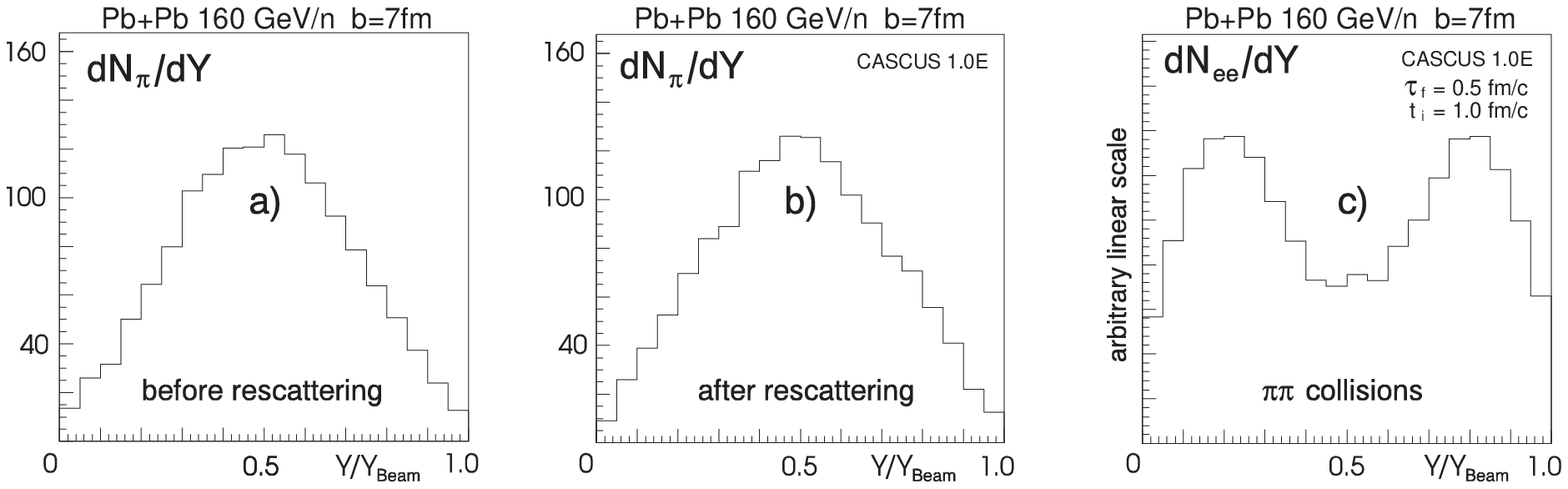}}
\vskip1.7pt
\centerline{\parbox{13cm} {\small {\bf Fig.3}
Rapidity distribution of dileptons c) and pions before/after rescattering. 
}}
\vskip0.15cm

The result shown on Fig.3 is obtained for the simulation of Pb+Pb
160 GeV/n events with values of rescattering model parameters
$\tau_f =0.5$ fm/c, $t_i=1.0$ fm/c. In this case number of collisions
per pion is close to 1.0 and minimum in the rapidity distribution of dileptons
is strongest. For higher collision rates
(obtained for smaller vaues of $\tau _f $ and $t_i$ parameters)
the minimum is significantly smaller. For number of collisions
per pion above 10 coll/$\pi$ the minimum disappeares at all. Such behaviour
indicates that the minimum in the rapidity distribution of dileptons
is generated by non-equilibrium
process in the expanding pion gas. This effect is studied in PhD
work \cite{PhD}.

From experimental point of view data on rapidity distribution
of dileptons produced in $p-A$ or $A-A$ collisions are rather
rare. It is not excluded that the only rapidity distribution
of dileptons which seem to originate from $\pi^+\pi^-$ annihilation
was obtained in pioneering experiments of
DLS collaboration at Bevalac accelerator in Berkeley \cite{DLS_min}.

\section{Azimuthal Asymmetry of Dileptons}\label{sec4}
Azimuthal asymmetries in transverse momentum distributions of 
particles have been
clearly identified in ultrarelativistic \cite{Ollie} non-central heavy ion
collisions.
This phenomenon is well understood as a consequence of collective
behaviour of nuclear matter or explained by absorption of secondary produced
particles in the spectator matter \cite{HGR}.
Recently also azimuthal asymmetries in
transverse momentum distributions of less abundant hadrons - $K$ mesons and
$\Lambda $ baryons \cite{QM_96_K_L} have been identified in HIC experiments.
However azimuthal asymmetries in transversal momentum distribution of
dileptons have been addressed neither experimentally nor theoretically
so far.

\vskip0.15cm
\centerline{\epsfxsize=5cm\epsffile{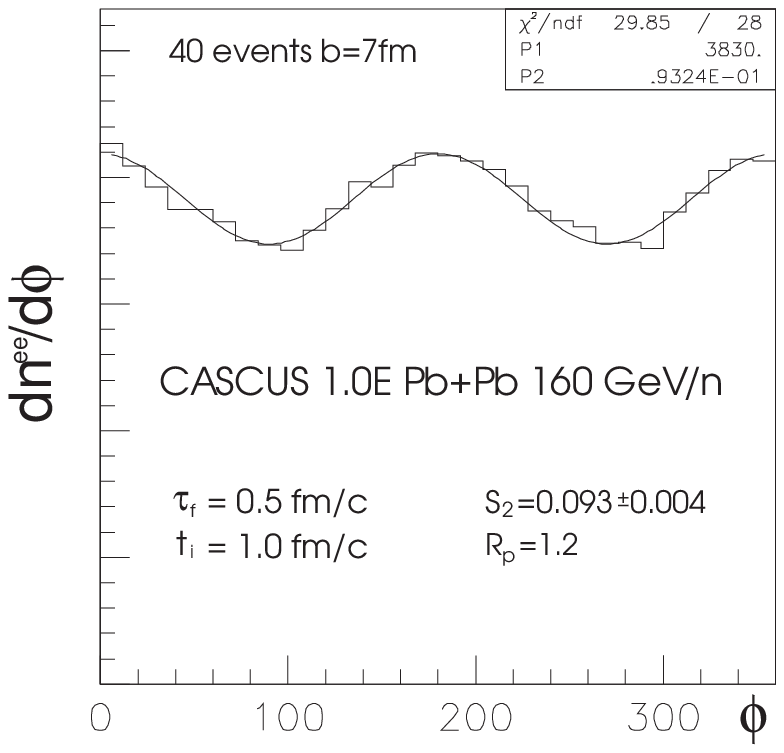}}
\vskip1.7pt
\centerline{\parbox{12.0cm} {\small {\bf Fig.4}
Asymmetry in transversal momentum distribution of dileptons.
}}
\vskip0.15cm

From theoretical point of view main mechanisms
generating azimuthal asymmetries in transversal momentum distributions of
hadrons are not applicable for dileptons. Dileptons leave freely
collision volume after being produced without any collective
final state interactions or absorption process.

In spite of this the computer simulation described shortly
in Section 2 predicts significant second-order
asymmetry in transversal momentum distribution of dilepton pairs.
Result shown in Fig.4 was obtained from the simulation of 40 non-central
Pb+Pb 160 GeV/n $b=7$ fm events. For each produced
dilepton pair orientation of momentum $\vec p_{ee} = \vec p_{e^+}+\vec p_{e^-}$
in respect to the reaction plane (impact parameter)
was determined. Histogram of azimuthal angles $\phi _{ee}$ of
dilepton pairs was filled and significant second order asymmetry was found.
The fit of the histogram shown in Fig.4 to the function:
\begin{equation}
R(\phi )=S_0[1+S_2\cdot \cos (2\phi )]
\label{S2}
\end{equation}
gave numerical value of the asymmetry coefficient $S_2=0.093\pm 0.004$.
Corresponding value of $R_p$ parameter (simplified for our purposes)
used to characterize squeeze-out
effect of hadrons at Bevalac energies \cite{Hans} is:
\begin{equation}
R_p=\frac{\langle p_y^2 \rangle }{\langle p_x^2 \rangle } = 1.202
\label{Rp}
\end{equation}

Orientation of the reaction plane is parallel to $y$ axis in our simulation,
azimuthal asymmetry of dileptons is oriented in the reaction plane.
From the analysis of second order asymmetry
in the transversal momentum distribution of pions
it comes out \cite{APS97,PhD} that the
asymmetry is strongest for the rate of collisions per pion close
to 4.0 coll/$\pi$.
For higher collision rates strength of the asymmetry decreases though it does
not tend to vanish in thermal equilibrium limit. 
The same behaviour is expected for the azimuthal
asymmetry of dileptons presented here for the first time.

Mechanism of generation of the azimuthal asymmetry of dileptons is
well understood \cite{PhD} at present. The asymmetry is closely related
to the spatial distribution of the pion gas in transversal plane which is
asymmetrical in non-central events.

\section{Summary}\label{sec5}
Two independent methods of the simulation of dilepton production from the
expanding pion gas were found to generate nearly identical results. 
Invariant mass spectrum of dileptons produced via $\pi^+\pi^-$ annihilation
channel was shown to be sensitive to momentum-space correlation in 
initial stage of the pion gas.

Generated rapidity distribution of dileptons exhibits unexpected
minimum at midrapidity. The minimum seems to be a consequence of 
a non-equilibrium process in the expanding pion gas since it becomes
weaker and disappears for higher number of collisions (above 10 coll/$\pi $)
in the pion gas.

For non-central collisions second order asymmetry in azimuthal distribution of
dileptons is found. This asymmetry is a consequence of asymmetrical overlapping
region of the colliding nuclei in transversal plane.
Azimuthal asymmetry of dileptons might be accessible experimentally if the
reaction plane orientation is determined from hadronic signal on dilepton
detector.
\section{Conclusions}\label{sec6}
In this contribution we study 
expansion of interacting pion 
gas created in ultrarelativistic heavy ion collisions.
Non-equilibrium and
asymmetrical features of the expansion
exhibit itself in rapidity and azimuthal distributions
of particles participating or created in the rescattering process.
Rescattering phenomenon is not related only to the
hadronic phase of nuclear matter produced in HIC. It can equally be present
also in prehadronic phase where it can lead \cite{Jupiter} to similar
effects as those revealed in the pion gas simulation. Investigation
of the non-equilibrium expansion of the parton gas possibly created also in
non-central collisions might shed light on other interesting aspects
\cite{Lourenco} of HIC research. 

\vskip0.7cm
\begin{center}
Content of this article was presented at HEP'97 conference in Jerusalem 
\cite{Jerusalem}.
\end{center}

\end{document}